\documentclass{article}

\usepackage{kmn,epsfig}

\begin{document}

\title[Radio-optical alignments]
{Radio-optical alignments in a low radio luminosity sample}

\author[Lacy et al.]{Mark Lacy$^1$, Susan E.\ Ridgway$^1$,  
Margrethe Wold$^2$, Per B.\ Lilje$^3$ \& Steve Rawlings$^1$\\
$^1$ Astrophysics, Department of Physics, Keble Road, Oxford, OX1 3RH,
U.K.\\
$^2$ Stockholm Observatory, S-133 36 Saltsj\"{o}baden, Sweden\\
$^3$ Institute of Theoretical Astrophysics, University of Oslo, P.O.\ Box
1029 Blindern, N-0315 Oslo, Norway}

\date{}

\pubyear{1998}

\maketitle

\begin{abstract}
We present an optically-based study of the alignment between
the radio axes and the optical major axes of eight $z\sim 0.7$ radio
galaxies in a 7C sample. The radio galaxies in this sample are
$\approx 20$-times less radio luminous than 3C galaxies at the same
redshift, and are significantly less radio-luminous than any other well-defined
samples studied to date. Using Nordic Optical Telescope images 
taken in good seeing conditions at rest-frame wavelengths just 
longward of the 4000\AA$\;$break, we find a statistically 
significant alignment effect in the 7C sample. Furthermore, in two
cases where the aligned components are well separated from the host 
we have been able to confirm spectroscopically that they are indeed 
at the same redshift as the radio galaxy. However, a quantitative 
analysis of the alignment in this sample and in a corresponding 3C sample 
from HST archival data indicates that the percentage of aligned flux may be
lower and of smaller spatial scale in the 7C sample.
Our study suggests that alignments on the 
50-kpc scale are probably closely related to the radio luminosity,
whereas those on the 15 kpc scale are not. We discuss these results in
the context of popular models for the alignment effect. 
\end{abstract}


\section{Introduction}

Studies of samples of
high radio luminosity radio galaxies at $z \geq  0.6$ 
show that the position angles of the optical and emission-line 
morphologies tend to 
align with the radio axis (McCarthy et al.\ 1987, Chambers et al.\ 1987).
High-resolution optical imaging with the Hubble Space Telescope (HST)
reveals that the aligned material is morphologically 
distributed in a wide variety of
ways (e.g. Best et al.\ 1996, 1997b), and is unlikely to be the result 
of a single mechanism. With the exceptions of 3C171 (Clark et al.\
1998) and 3C368 (Dickson et al.\ 1995;
Stockton, Ridgway \& Kellogg 1996), nebular continuum
fails to explain more than a fraction of the aligned light
in objects for which deep spectroscopy has been obtained. Scattering
of a hidden quasar nucleus is expected to add to the aligned light,
and the detection of broad, polarised Mg{\sc ii} emission in a number of
these objects (e.g.\ 3C265 and 3C324;
Dey \& Spinrad 1996; Cimatti et al.\ 1996) shows that this can be
an important contributor.  The closely aligned morphologies
seen in the WFPC2 imaging are, however, inconsistent with pure scattering
models, and star formation induced by the passage of the radio jet
has been proposed by many (e.g.\ Chambers et al. 1990, Best et al. 1996, 
Dey et al. 1997).

Most proposed mechanisms for producing radio -- optical alignments 
are expected to depend to some extent on the AGN luminosity. 
The radio luminosity and 
the strength of narrow optical emission lines in radio galaxies
have been known to correlate for some time (Baum \& Heckman 1989; 
Rawlings et al.\ 1989). 
Recently, a correlation of the optical continuum luminosity of steep-spectrum 
radio-loud quasars with radio luminosity has also been established
(Serjeant et al.\ 1998). Both these imply a close relationship between 
the optical/UV luminosity of the AGN and the radio luminosity, presumably
via a correlation with the bulk kinetic power of the radio jets 
(Rawlings \& Saunders 1991). Given this, it is not surprising that the
optical emission aligned with the radio jet seen in high redshift
radio galaxies might also correlate with radio luminosity.
Two studies in the near-infrared, that of Dunlop \& Peacock (1993) and 
of Eales et al.\ (1997) found that this ``alignment effect'' was unmeasurably
small in samples of $z\sim 1$ radio sources approximately ten and four times 
fainter than 3C respectively. Both these studies were, however, 
carried out under conditions of poor 
($\stackrel{>}{_{\sim}}1^{''}$) seeing, and in the $K$-band (roughly
1-micron in the rest-frame), where the 
alignment effect is weak even in 3C (Rigler et al.\ 1992). They therefore
could not usefully constrain how strongly the alignment effect depended 
on radio luminosity. 

The question of the luminosity dependence of the alignment effect 
is also related to the extent to which the onset of the
alignment effect at $z\sim 0.6$ in flux-limited samples such as 3C is
a true evolutionary effect, or just an effect due to sources in a
flux-limited sample becoming more luminous with redshift. 

In this paper, we present optical images of a sample of $z\sim 0.7$
radio galaxies whose radio luminosity is approximately twenty times 
lower than those of 3C radio galaxies at the same redshift. Using these
images and data on a comparison sample of 3C objects obtained from the 
HST archive, we quantify and compare the strength of the alignment effect 
in the two samples. We assume an 
$H_{0}$ = 50 km s$^{-1}$ Mpc$^{-1}$, $q_{0}$ = 0.5 cosmology throughout.

\section{Observations and Data Reduction}

\subsection{Imaging}

Images of a sub-sample of the objects in a low-frequency selected sample
of radio sources in the North Ecliptic Cap (NEC) were obtained on the 
Nordic Optical Telescope (NOT) in July 1995 and May 1996 (Table 1). 
The sub-sample
consisted of all but one (missed out at random) of the nine extended
(radio size $\theta_{\rm r}>1$\arcsec\ ) $0.5<z<0.82$ 
radio galaxies in a sample based on that of Lacy et al.\ (1993), with 
a few additions and deletions made to make the sample complete to a flux
density of 0.5 Jy at 151 MHz (the ``7C-III'' sample; Lacy et al.\ 1998a).
Details of the properties of the sample are given in Table 2.
The 1995 July observations were made with the BROCAM camera with a 
TEK 1024 square CCD and a 0\farcs176 pixel$^{-1}$ scale. 
Those in 1996 May were 
made with the HIRAC camera, a 2048 square Loral CCD and a 0\farcs108
pixel$^{-1}$ scale. Images were made in either $R$ or $I$, with $0.5<z<0.6$ 
objects in $R$ and $0.6<z<0.82$ in $I$, sampling the rest frame 
SED at about 450nm, just above the 4000\AA$\;$break. The images are
shown in Fig.\ 1. 
\begin{figure*}[h!t]
\begin{picture}(500,500)
\put(50,0){\includegraphics[scale=0.7,angle=0]{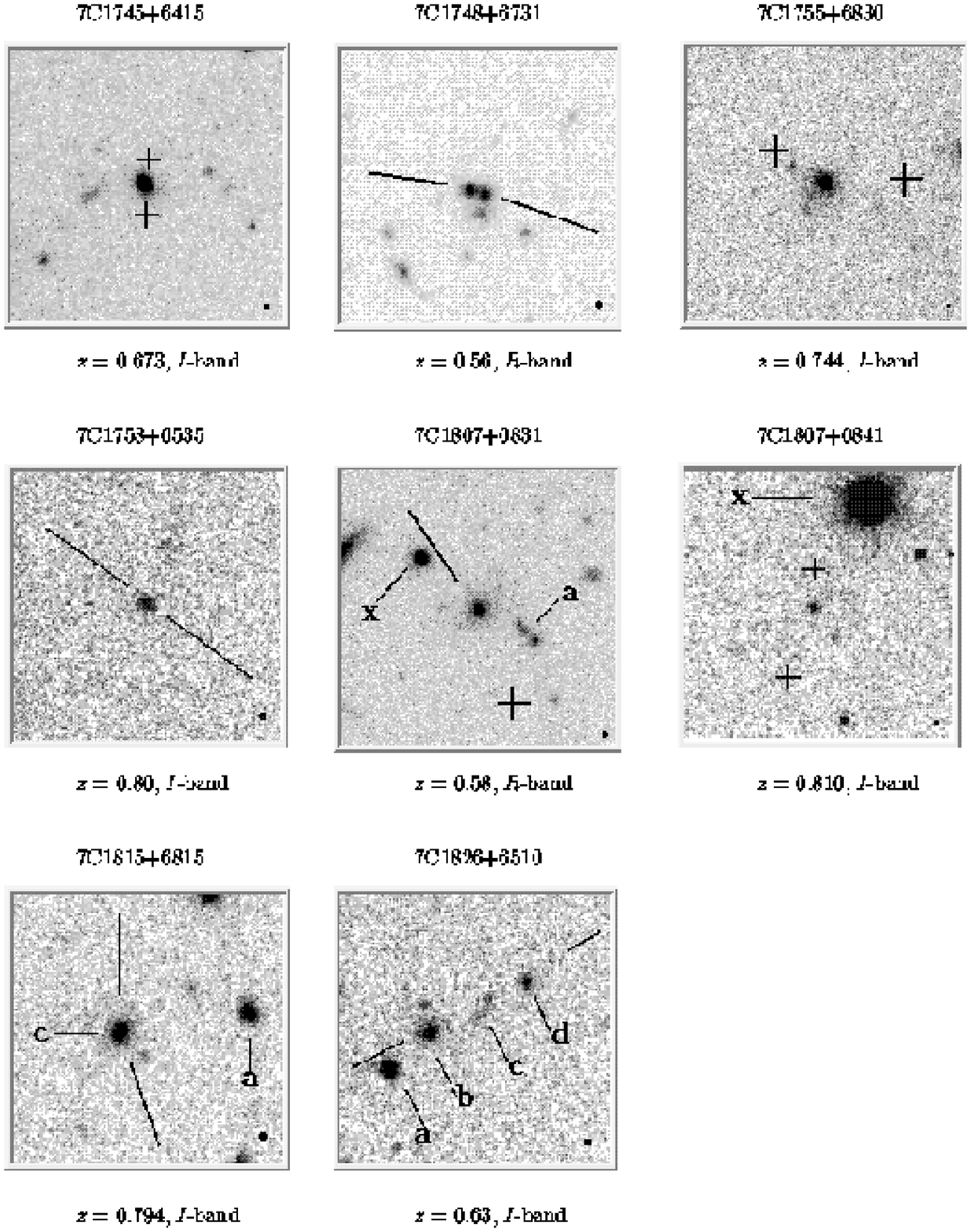}}
\end{picture}
\caption{The 7C galaxies analyzed in this paper, with N up, E left. The images
are 28\arcsec\ square. The black crosses designate the
positions of radio hotspots, while the black lines show directions to the
hot spots if they fall outside the field. Multiple objects have
their individual sub-components labelled. Foreground stars removed 
from the aperture analysis are labelled with an `X' [apart from 
7C1826+6510a which was removed but is labelled `a' for consistency 
with Lacy et al.\ (1998b)]. The dot in the SW
corner of the image indicates the FWHM of the seeing disc.}
\end{figure*}

\subsection{Spectroscopy}
Optical spectra of most of the objects in the 7C-III sample were obtained
with the ISIS spectrograph on the 
William Herschel Telescope (WHT) on 1995 July 28 - August 1. Both the red and
blue arms were used, with a 570nm dichroic used to split the beams and
TEK CCD detectors in both arms. The R300R and R300B gratings were used
on the red and blue arms respectively.
7C 1815+6815 was observed on 1993 June 17 with a similar set-up, but
an EEV CCD on the red arm with a 540nm dichroic. A slit width of
3$^{''}$ was used, which gave a velocity resolution of 
$\approx 1500$kms$^{-1}$ for an object filling the slit. Full details of 
the observations and data reduction 
will be given in Lacy et al.\ (1998c). Emission line fluxes 
are listed in Table 2, where the estimated emission line contamination of the 
optical images is also shown.

\begin{figure*}
\begin{picture}(500,650)
\put(0,0){\includegraphics[scale=1.0]{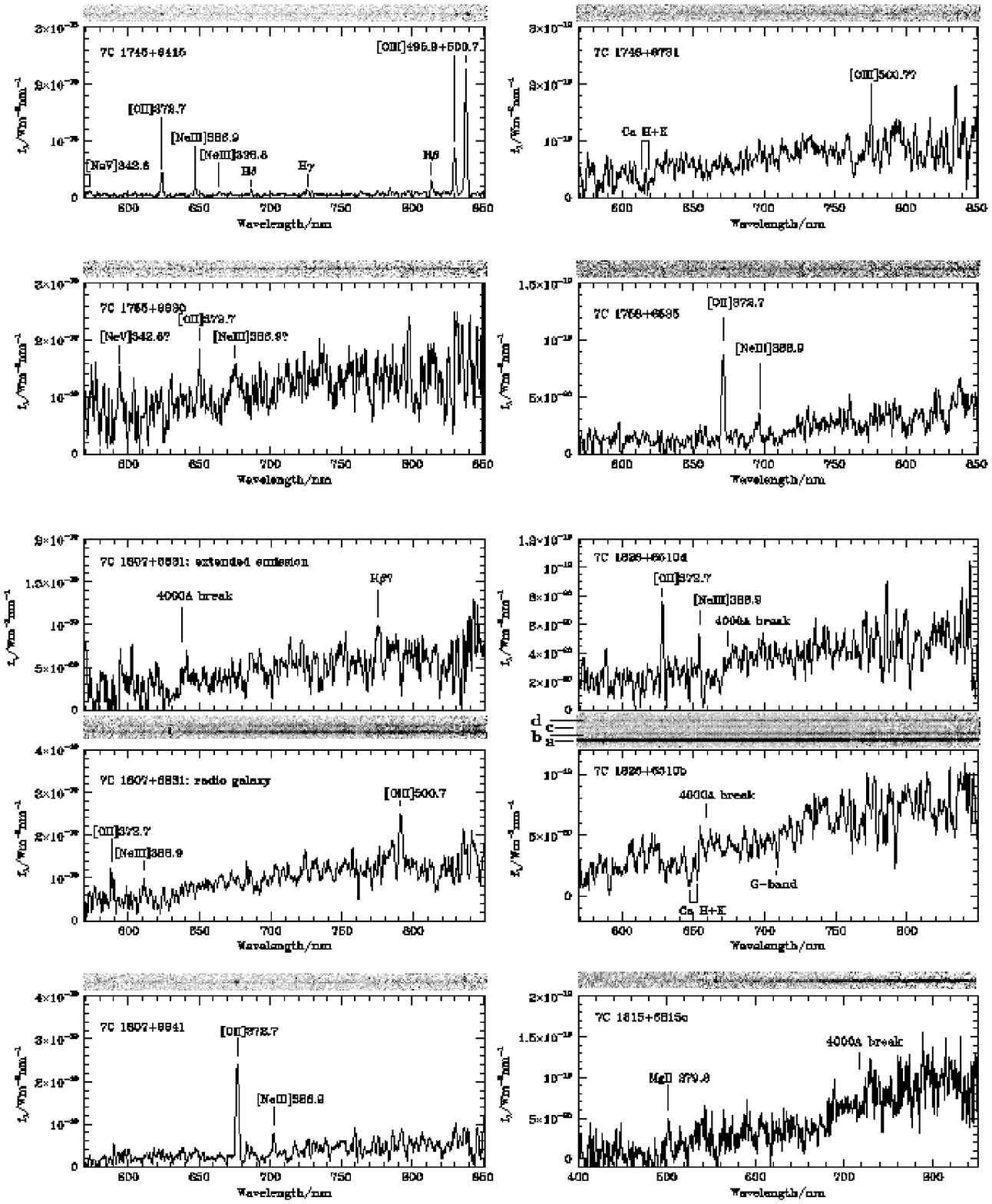}}
\end{picture}
\caption{Spectra of the 7C objects. The spectra have been smoothed with 
a 1.5 nm (5-pixel) box-car filter. Greyscales of the 2D spectra are also
shown.} 

\end{figure*}

\begin{table*}
\caption{Observations of the low luminosity sample}
\begin{tabular}{llllllllr}
Name &R.A.\ (1950)& Dec.\ (1950)& NOT imaging & seeing &$t_{\rm exp}$& WHT spectrum & $t_{\rm exp}$& slit PA\\
     &            &             &date \& filter&/arcsec&/min         & date 
     & /min        &         \\
7C 1745+6415&17 45 07.30&+64 15 28.6& 23/07/96 $I$ & 0.6 & 40& 31/07/95 & 9&177\\
7C 1748+6731&17 48 50.70&+67 31 38.8& 11/05/97 $R$ & 0.8 & 40& 29/07/95 &10&0\\
7C 1755+6830&17 55 54.94&+68 30 55.9& 10/05/97 $I$ & 0.5 & 40& 31/07/95 & 5&65\\
7C 1758+6535&17 58 11.58&+65 35 17.1& 24/07/96 $I$ & 0.9 & 40& 29/07/95 &30&52\\  
7C 1807+6831&18 07 06.05&+68 31 11.5& 14/05/97 $R$ & 0.7 & 40& 29/07/95 &10&60\\
7C 1807+6841&18 07 47.32&+68 41 20.4& 25/07/96 $I$ & 0.7 & 40& 28/07/95 &26&165\\
7C 1815+6815&18 15 45.69&+68 17 19.2& 24/07/96 $I$ & 0.9 & 40& 17/06/93 &28&100\\
7C 1826+6510&18 26 31.31&+65 10 45.5& 25/07/97 $I$ & 0.8 & 70& 28/07/95 &30&123\\
\end{tabular}

\noindent
Note: positions are those of the optical identification, and should be accurate
to within $\approx 1$ arcsec.
\end{table*}

\begin{table*}
\caption{Properties of the low luminosity sample}
\begin{tabular}{lllrlrrl}
Name &$S_{151}$&$z$&Radio&Emission&Flux& percent line & Notes\\
     &/Jy &&size/arcsec&line &/$10^{-20}$&contamination\ &\\
     &    &&            &     &Wm$^{-2}$  &        &\\
7C 1745+6415&0.59 & 0.673&5.6&[O{\sc ii}]372.7&80&40&\\
            &     &      &&H$\beta$       & 40   &&\\
            &     &      &&[O{\sc iii}]495.9&140 &&\\
            &     &      &&[O{\sc iii}]500.7&400 &&\\
7C 1748+6731&0.64 & 0.56 &108&[O{\sc iii}]500.7?&9&$<1$&absorption line $z$\\
7C 1755+6830&1.52 & 0.744&8.9&[O{\sc ii}]372.7&20&2& \\
            &     &      &   &[Ne{\sc iii}]386.9&17&&\\
            &     &      &   &[Ne{\sc v}]342.6? &20&&\\
7C 1758+6535&1.13 & 0.80 &106&[O{\sc ii}]372.7&18&1&\\  
            &     &      &   &[Ne{\sc iii}]386.9&10&&\\
7C 1807+6831&2.12 & 0.580&29 &[O{\sc ii}]372.7&11&& \\
            &     &      &   &[Ne{\sc iii}]386.9& 8&&\\
            &     &      &   &[O{\sc iii}]500.7& 22&1&\\
7C 1807+6841&0.60 & 0.816&12 &[O{\sc ii}]372.7  & 52&$<1$&\\
            &     &      &   &[Ne{\sc iii}]386.9& 11&&\\
7C 1815+6815&1.37 & 0.794&200&Mg{\sc ii} 279.8&20&$<1$&plus 4000\AA$\;$break\\
7C 1826+6510&1.39 & 0.646& 34&                &  &$<1$&absorption line $z$\\
\end{tabular}

\noindent
Notes: in all cases the linewidths were below the instrumental resolution 
for an object filling the slit ($\approx$ 1500 kms$^{-1}$) so have not been 
listed. The errors on the line fluxes are $\approx 15$ per cent, excluding
aperture corrections. Percentage 
emission line contamination is the amount of contamination
in the filter of observation from the emission lines listed which fall in
the filter bandpass. In a few cases emission lines which fall in the $I$-band
are beyond the end of the spectra; in these cases their fluxes have been 
estimated using the emission line strengths in McCarthy (1993).
\end{table*}

\subsection{Archival HST data}

Data on 11 radio galaxies in the sample of Laing, Riley \& Longair
(1983; hereafter LRL) 
in the redshift range $0.5<z<0.82$, with  $\theta_{\rm r}>1^{''}$
and Dec. $<60^{\circ}$  were obtained from the HST archive. This sample is 
complete within the selection criteria with the following exceptions: 
3C225B was
excluded from the sample due to an uncertainty in its optical identification,
3C352 was excluded due to the presence of a nearby bright star, and 3C55 was 
inadvertently
missed out due to its redshift in LRL being incorrect. The exclusion of these
objects is very unlikely to bias the results in any way.
Details of the observations are given in Table 3.

\begin{table}
\caption{HST observations of the 3C sample}
\begin{tabular}{llcclr}
Name & $z$ & Date &  Filter & CCD & $t_{\rm exp}$ \\
     &     &         &      &     &     \\
3C 34& 0.69& 10/06/94&F785LP& WF3 & 1700\\
3C 41& 0.794&29/07/94&F785LP& WF3 & 1700\\
3C 172&0.52 &19/04/04&F702W & PC  & 600\\
3C 226& 0.818&04/05/94&F785LP&WF3 &1700\\
3C 228& 0.552& 14/03/94&F702W&PC  & 300\\
3C 247& 0.75 & 29/03/96&F814W&WF3 &2400\\
3C 265&0.811 & 29/05/94&F785LP&WF3&1700\\ 
3C 277.2&0.766&20/06/96&F814W&WF3 &2400\\
3C 337&0.63  & 24/08/95&F814W &WF3&1400\\
3C 340&0.76  &25/04/94 &F785LP&WF3&1700\\
3C 441&0.708 &30/05/94&F785LP&WF3&1700\\ 
\end{tabular}
\end{table}

\section{The Alignment Effect in the 7C sample} 

A visual inspection of the images (Fig.\ 1) reveals a number of objects 
in which the optical morphology appears aligned with the radio axis. The
morphologies of the aligned material
differ: in two cases, the hosts are apparently ellipticals
with the major axis within 30 deg.\ of the radio PA, while others
(e.g.\ 7C 1748+6731 and 7C 1826+6510)
have multicomponent morphologies along the 
radio axis similar to that seen in the high luminosity 
3C sources. Some individual cases are discussed in more detail below.

\paragraph*{7C 1745+6415}

This radio galaxy looks like a smooth elliptical, but in fact about 40 
percent of the $I$-band light is from emission lines, and
consequently the morphology visible in the $I$-band image
may be strongly affected by nebular emission. It is therefore 
not clear in this case whether or not the stellar continuum light is aligned.

\paragraph*{7C 1748+6731}

This is the most well-aligned object in either sample due to its
double structure which is aligned almost perfectly with the radio
axis. 

\paragraph*{7C 1807+6831}
The peculiar double component `a' seen 
to the SW of the host galaxy has the spectrum of a red stellar
population consistent with the host redshift. The aligned component has 
a tentative H$\beta$ emission line at $z=0.595$, redshifted by $2800$
kms$^{-1}$ relative to the host galaxy. This was checked by 
cross-correlating the spectra of the host and extended emission (after
subtraction of emission lines)
with a template spectrum obtained from a high signal-to-noise observation
of a low redshift member of the 7C-III sample. This cross-correlation
technique yields a redshift of the aligned component of 
$1300 \pm 700$ kms$^{-1}$ with respect to the host, broadly
consistent within the errors. 

\paragraph*{7C 1815+6815}

In Lacy et al.\ (1993) two possible identifications for this large
angular size source
were proposed. We have adopted `c' as the identification as it has a
weak emission line and is brighter in the $H$-band (Lacy et al.\
1998b). The alternative identification, `a' has a similar stellar 
spectrum to `c' and is probably in the same group or cluster.

\paragraph*{7C 1826+6510}

The radio central component detected in an A-array 8 GHz VLA observation (Lacy
et al.\ 1998b) shows that
component `b' of Lacy et al.\ (1998b) 
is the host galaxy (as claimed by Lacy et al.\ 1993), although only the 
aligned component `d' has detected emission lines. Both `b' and `d' have 
spectra dominated by old stars (Fig.\ 2). The diffuse component `c' which lies
between `b' and `d' also has a red SED, but is too faint for us to be
able to tell if it contains stars, or to measure a redshift for. 
The redshifts of `b' and the stellar
component of `d' were determined by the cross-correlation technique 
described above. We find that `b' has a
redshift of $0.646\pm 0.001$ and `d' a redshift of $0.685 \pm 0.001$
(agreeing with the emission line redshift). The velocity difference 
between these two components is $7100 \pm 400$ kms$^{-1}$, which is
high, but not implausible if they are members of a rich
cluster. Although clearly well-aligned, `d' lies outside our largest
aperture so is excluded from our analysis in the next section.
Component `a' is a foreground star and was removed prior to the image 
analysis.

\section{Quantifying the alignment effect}

To test whether our apparent detection of an alignment effect in the 
7C-{\sc iii} sample is statistically significant, 
we have attempted to make a quantitative
measurement of the ``alignment effect'' in each galaxy
and compared these measurements to those derived for the 
sample of 3C galaxies in Table 3. 

We have used two methods to measure the alignment of the galaxies in
these samples: first, a standard moment analysis to derive the difference
in the
optical position angle from the radio position angle ($\Delta$ PA)
(Rigler et al. 1992, Dunlop \& Peacock 1993, Ridgway \& Stockton 1997), 
and second, 
a simple derivation of the percentage aligned flux, which should be 
less susceptible to skewing by companions and differences in observational
resolution. To derive this percentage aligned flux, 
we have summed the flux within $\pm 45^{\circ}$ 
of the lines joining the nucleus to each of the radio hotspots,
and subtracted the remaining flux. We exclude from the sums the central portion
of each galaxy within the FWHM of the seeing for the
ground-based data, and for the archival HST data, we exclude a circular region 
with a diameter (0\farcs5) comparable to the FWHM of the best seeing
ground-based images. 
This percentage aligned flux value will be positive for images with more flux
within the cone of the radio axes than without,
and will be negative for objects with more counter-aligned than aligned flux.  

To make intercomparisons between objects in this fairly inhomeogenous dataset,
we must account both for differences in the depth of the observations and
the spread in redshift across the samples. Before analysis, 
we have therefore resampled 
the 7C NOT and 3C WFPC2 
images to a common pixel scale of 0\farcs025 pixel$^{-1}$, 
and then corrected the surface brightnesses
for the differences in cosmological dimming by normalizing
to the median redshift of the samples ($z$=0.7) (by assuming that $F_{\lambda}
\propto (1 + z)^{-5}$, i.e.\ that the intrinsic spectra of the galaxies are 
approximately flat in $F_{\lambda}$ at these wavelengths). 

A well-known problem with the standard position angle analysis is the
subjectivity of choosing the isophotal cutoff level and the aperture within
which to calculate the moments (Dunlop \& Peacock 1993). 
Ridgway \& Stockton (1997) show that
in many cases, however, there is a ``dominant'' position angle which 
is approximately constant over a wide range of isophotal cutoffs. The
choice of aperture is perhaps more critical, as emission at large
radii can easily skew the position angles due to the fact that the PA
is determined from the second moment of the intensity distribution. 
Thus in large apertures, any radio--optical alignments tend to be 
dominated by the contribution of extended emission and companion 
galaxies rather than any alignment of the host galaxy itself.

This is borne out in Fig.\ 3 (a) and (c)  
where the results of the PA analysis are plotted as a
function of isophotal level and aperture size. Provided the isophotes
are high enough to be clear of the noise and low enough to include
an area of several resolution elements above the isophotal cutoff,
the value of the PA in a given small ($\stackrel{<}{_{\sim}}$
3\arcsec\ ) aperture is independent 
(to about $\pm 10^{\circ}$) of the exact isophotal level. As discussed
above the larger apertures are more sensitive to the isophotal cutoff
as they tend to include low surface brightness emission and faint
companions which disappear as the isophotal cutoff is raised.

For most of the analysis in this
paper, we have chosen a single isophotal cutoff, used for all the images, of 
$1.0 \times 10^{-23}$ Wm$^{-2}$nm$^{-1}$ pixel$^{-1}$,
about 2.5 times the median normalized sky sigma 
in these frames, for both the position angle analysis and
for the percentage aligned flux analysis. We chose our apertures based
on inspection of the behavior of the integrated percentage 
aligned flux in increasing apertures up to about 10\arcsec\ for
objects from both samples [e.g.\ Fig.\ 3 (b) and (d)]. These indicated 
that in general there were two spatial regions which tended to
contain the major flux contributions:
within about 2\arcsec\ ($\sim$15 kpc) and within about 6\arcsec\
($\sim$50 kpc). These corresponded physically to the approximate
isophotal extents of the host galaxies, and to the distance to the 
nearest few companions respectively. Thus although our choice of
apertures is to some extent arbitary, they correspond to reasonable
physical length scales and, as Fig.\ 3 indicates, varying these
apertures by factors of 0.7-1.5 will make little difference to the
results. Thus we chose 15 kpc and 50 kpc as the radii of
our standard apertures for our analyses.

\begin{figure*}
\begin{picture}(500,500)
\put(0,0){\includegraphics{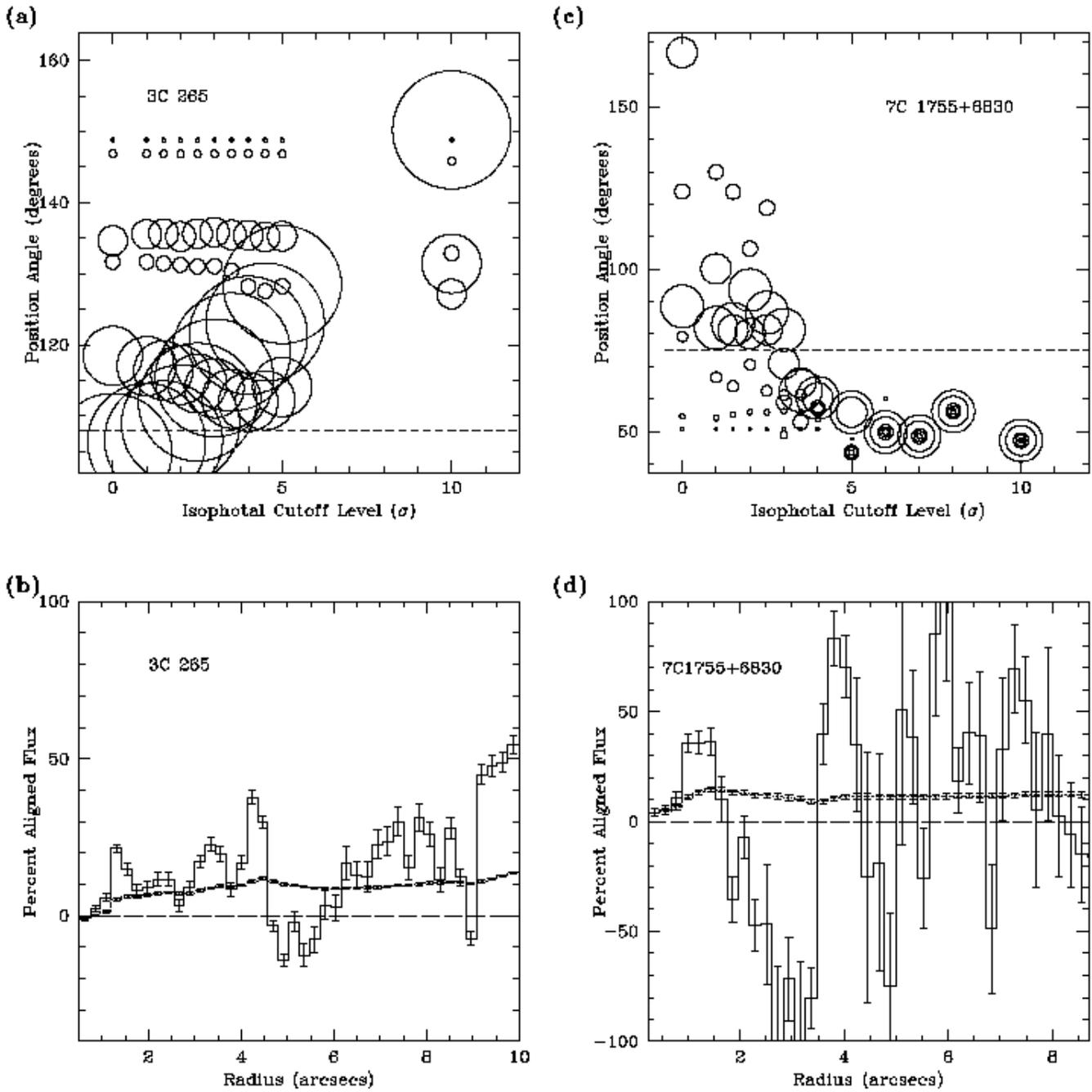}}
\end{picture}
\caption{Two examples of results from the analysis methods
discussed in this paper. First, for the well-known object 3C265, 
(a) position angle from a moment analysis versus isophotal level in
units of the noise level on the image, $\sigma$. This plot shows how the 
position angle measured in a given aperture varies as the isophotal cutoff is 
changed. The position angle, measured 
in apertures of radius 0\farcs3, 0\farcs6, 1\farcs2, 2\farcs4, 4\farcs8 and
9\farcs6 is plotted with circles whose sizes are proportional to the 
aperture radius. The dashed line is plotted at the PA of the radio axis. In
(b) the percentage aligned flux versus aperture radius for 3C265 is plotted. 
The solid line is the percentage aligned flux within annuli whose radial 
width is the width of the histogram bar and 
the dotted line the cumulative percentage aligned flux as a function 
of radius, obtained by summing the contributions from each annulus within 
that radius. Second, (c) and (d) are the same plots for the fairly typical 7C 
radio galaxy 7C 1755+6830(c) and (d). The aperture radii are 
0\farcs54, 1\farcs1, 2\farcs2, 3\farcs0, 6\farcs2 and 8\farcs6.}
\end{figure*}

In all these analyses, we have used as the 
center the optical center of the identification of the host galaxy.
For 7C 1748+6731, we have used the halfway point between the two symmetric
optical components (and in this case, also used an exclusion
radius of only 0\farcs05 for the flux analysis). 

A number of objects had foreground stars within the larger
aperture which were masked out from the analysis. Those for objects 
in the 7C sample are marked in Fig.\ 1. In the 3C sample only 3C 247
had a star removed; the object is to the NE of the radio galaxy and 
can be recognised from diffraction spikes 
visible in figure 16(a) of Best et al.\ (1997b).

For each object in the sample we have calculated the $\Delta$PA and the
percentage aligned flux as described above, within these two apertures after 
making the isophotal cutoff. For many of these objects,
which were observed longer or were at lower than median redshift,
this standard isophotal cutoff may result in losing obviously well-detected
low surface brightness material: for example, large scale extended aligned
material in 3C 277.2 falls below this cutoff. This is necessary, of course,
to make the comparison on an unbiassed basis. 
We have also calculated
the percentage aligned flux in the annulus between 15 kpc and 50 kpc,
and we give in Table 4 the results of these analyses for both samples. 
Also given are the unnormalized aperture magnitudes for the objects.

As a check for the robustness of our result to varying the isophotal levels
the PA and percentage flux analysis was repeated, but choosing the isophotal
level to be $2.5 \sigma$ above the background of each individual image.
In most cases this made no difference to the PAs and 
percentage aligned fluxes, and although a couple of individual results
changed (for example that for 3C277.2 discussed above) the effect on
the statistical analysis discussed in the next section was
negligible. 

\begin{table*}
\caption{Results of alignment analysis}
{\small
\begin{tabular}{lccccccc}
Name &\% Aligned &\% Aligned& \% Aligned & $\Delta$PA & $\Delta$PA &
m(AB) & m(AB)\\
     &   Flux   &  Flux    &  Flux   &           &           &       &\\
     & (15 kpc) & (50 kpc) & (15 -- 50 kpc) &  (15 kpc) & (50 kpc)& (15kpc) & (50 kpc)\\
&       &       &       &       &       &       &\\
3C34   & 4.1 $\pm$  1 &$-$44 $\pm$ 1 & $-$68.1 $\pm$ 1 & 20 & 73 &
20.0 & 19.7 \\
3C41   & $-$6.9 $\pm$ 1 &  16 $\pm$ 1 & 96.3 $\pm$ 3 &  88 & 22 & 20.9
& 20.4 \\
3C172  & 10.6 $\pm$ 1 & 37.2 $\pm$ 1 & 44.8 $\pm$ 1 & 12 & 23 & 19.9 & 18.9\\
3C226  & 24.8 $\pm$  1 &   17.1 $\pm$ 1 & $-$8.5 $\pm$ 2 &  2.7 & 36 &
20.9 & 20.3\\
3C228  & $-$9.9 $\pm$  2 &  - & - & 62 & - & 19.8 & - \\
3C247  & $-$7.6 $\pm$ 1 & 41.0 $\pm$ 1 & 89.6 $\pm$ 1 &  74& 14 &20.2
& 19.1  \\
3C265  & 7.8  $\pm$ 1 &   15.4 $\pm$ 1 & 23.5 $\pm$ 1 &  19 &11  & 20.1 & 19.3 \\
3C277.2 & 22.1 $\pm$  1 & 3.2 $\pm$ 1 & $-$90.0 $\pm$ 1 & 11 &  75 & 20.6 & 20.2\\
3C337  & 25.2 $\pm$  1 & 55.4 $\pm$ 1 & 79.7 $\pm$ 1 & 2.5 & 10 & 20.7 & 20.2\\
3C340  & 15.8 $\pm$  1 &  19.4  $\pm$ 1  & 92.9 $\pm$ 7 & 13 & 18 & 21.1 & 20.8\\
3C441  & 18.6 $\pm$  1 &  27.5$\pm$ 1 & 33.8 $\pm$ 1 &  17 & 7 & 20.8 & 19.9\\
Median 3C&      11  & 18 & 39   & 17   & 20 & 20.6 & 20.1 \\

  &              &              &      & &     &     &    \\
7C1745+6415&  6.5 $\pm$ 1 & 6.5 $\pm$ 1 & 0 $\pm$ 0 & 35 & 35 & 19.7 & 19.5\\
7C1748+6731&  100 $\pm$  2& 100 $\pm$ 2 &  0 $\pm$ 0 & 1.5 & 1.5 & 21.1 & 20.4 \\
7C1755+6830&  12.2 $\pm$   1&  8.8 $\pm$ 1 &$-$6.1 $\pm$ 3 &  22 & 4  & 21.0 & 20.5\\
7C1758+6535&   $-$3.7 $\pm$   7&  $-$13.4 $\pm$ 6 & $-$47 $\pm$ 15&  2 & 72 & 21.4 & 20.9 \\
7C1807+6831&   15.1 $\pm$ 1&   24.4 $\pm$ 1   & 81 $\pm$ 4 & 11 & 35 & 20.6 & 19.9 \\
7C1807+6841&   $-$9.2 $\pm$   3&   $-$11.5 $\pm$ 2 & $-$12.3 $\pm$ 2 & 17 & 70 &21.6 & -    \\
7C1815+6815&  6.8 $\pm$ 2&   7.3 $\pm$ 2 & 16.9 $\pm$ 10 & 28 & 25 & 20.2 & 19.7 \\
7C1826+6510&   10.3 $\pm$   3&    4.5 $\pm$ 3   &$-$11.9 $\pm$ 6 &  11 & 22 & 20.1 & 19.5 \\
 Median 7C  &9          & 7  & $-$6         &  14 & 30 & 20.8 & 19.9 \\
\end{tabular}
}

\footnotesize
(1) The errors given are the statistical errors based on the contribution of
sky noise to each sum.
(2) The ``0 $\pm$ 0'' values mean that no flux was left above
the isophotal cutoff in those annuli.
(3) The 3C 228 image was a very short single PC exposure; only the inner portion
was used since all CRs were removed by hand.
\end{table*}

\section{Results}

The results of our analysis are summarised in Table 4 and Fig.\ 4. 
We find that the median percentage aligned flux in the both the 7C and
3C radio sources is positive within the smaller (15 kpc) aperture. The
dispersion in the percentage aligned flux is large for both samples,
making a meaningful statistical analysis difficult with these small samples.
Nevertheless we have attempted to use simple non-parametic statistical
methods to estimate the significance of our results.

We give in Figures 4 and 5 histograms of the results of 
the $\Delta$PA and percentage
aligned flux analysis within the 15 and 50 kpc apertures for the 7C and 3C
samples. The percentage aligned flux median values given in
Table 4 seem to indicate a comparable degree of alignment in the 7C as in
the 3C sample, at least at the 15 kpc radius.
In the region between 15 kpc and 50 kpc,
however, the 3C sample seems more aligned.

To determine whether these results indicate a significant alignment
in the 7C and 3C samples, we have made a number of statistical tests,
the results of which we give in Table 5. 
First we determine whether the $\Delta$PA values are consistent
with random orientation; a Kolmogorov-Smirnov (K.S.) test 
of the values versus a uniform distribution (Table 5, lines 1 \& 2)
shows
that at 15 kpc both the 7C and the 3C sample have a probability of 
less than 1 per cent of being randomly oriented.
The same test for the 50 kpc aperture is less conclusive for both
samples, as might be expected from an increased slewing of the PA by 
companion objects, but particularly for the 7C sample (line 3).
What about the percentage aligned flux values? 
We have made a K.S. test to determine whether the 7C and 3C percentage aligned
flux distributions differ significantly, and find that in the 15 kpc aperture,
there is no significant difference (Table 5, line 5).
Within the 50 kpc aperture, and particularly in the 15 -- 50 kpc annulus,
we find that the probabilities of having such different distributions 
by chance is significantly reduced (Table 5, lines 6,7,8).

From these tests we can conclude that at least at the 15 kpc scale the 7C
sample we have observed seems to exhibit a significant alignment effect. 
At the larger 50 kpc scale, while the 3C sample continues to show a similar
or perhaps increasing amount of aligned flux, the 7C sample is less well
aligned.

One potential source of bias in this study is the difference in quasar 
fraction between the 3C and 7C samples: in the 
3C sample of LRL 13/32 of all the radio sources in the redshift range 
$0.5\leq z <0.82$ are quasars, compared to 0/10 objects in the total 
7C-{\sc iii} sample. If orientation-based unified schemes of quasars and 
radio galaxies are true, this may mean that we are missing some 
3C objects which are close to the line of sight by selecting only 
radio galaxies, and hence the mean projected size of
the aligned structures in the 3C galaxies will be overestimated. The 
difference in quasar
fraction may be due to the 7C radio galaxies being members of the class of weak
emission line FRII radio galaxies identified by Laing et al.\ (1994) which
have only low optical luminosity quasar nuclei, if any. The quasar fraction
in LRL in our chosen redshift range corresponds to an angle to the line of 
sight of 54$^{\circ}$ being the 
angle within which a quasar is seen rather than a radio galaxy. On this basis
we might expect the mean projected size of the aligned regions in the 3C 
sample to be 20 percent larger than that for an isotropic 
population, but this cannot on its own account for the difference in the 
scale sizes of the 7C and 3C aligned regions.

\subsection{Correlation with overall radio spectral index}

Dunlop \& Peacock (1993) claim to find a correlation of the fraction
of blue light with a combination of radio power and spectral index,
which they argue is linked to the aligned light. We
find no correlation between aligned flux and rest-frame
spectral index measured at 1 GHz in either the 3C or 7C samples, 
though our samples are somewhat smaller. The lack of any
correlation would suggest that the blue light and the aligned light
were not simply related: this certainly seems to be the case for the
red aligned components discussed in Section 6.4.

\subsection{Correlation with radio source size}

Best et al.\ (1996) note a tendency for small $z\sim 1$ radio sources to have 
very well-aligned, knotty structures, whereas the larger sources often have
bimodal, generally less well-aligned structures. We see no 
tendency in either the 3C or the 7C samples for either the 
difference in position angle or the aligned flux to correlate with 
size either at 15 or 50 kpc, although we note that our samples are of both 
lower mean radio luminosity and lower mean redshift 
than the Best et al.\ (1996) objects.

\begin{figure*}
\includegraphics[scale=0.75,angle=0]{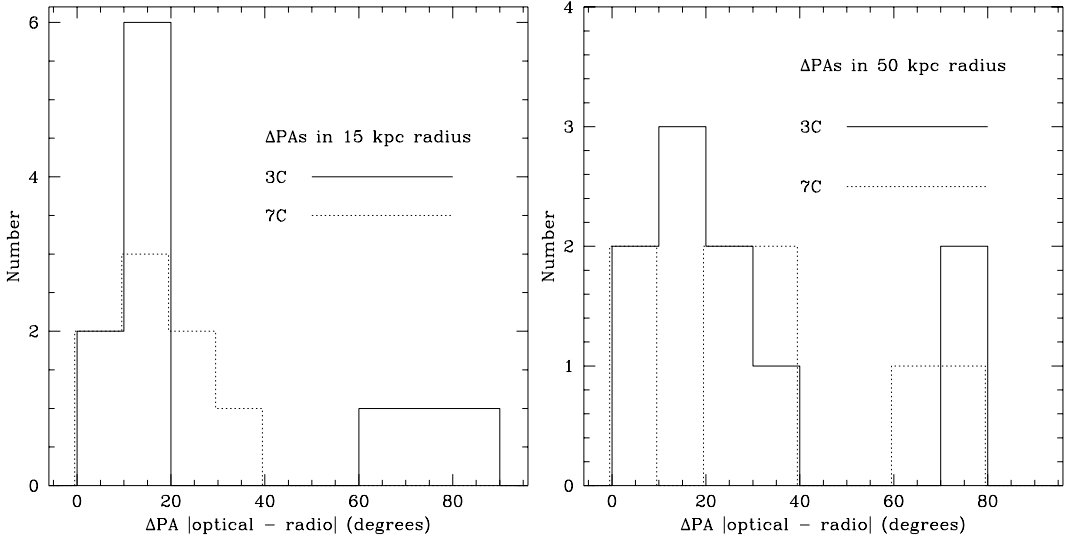}
\caption{Left: $\Delta$PAs (difference between the optical and radio axes)
for the 3C and 7C galaxies within the 15 kpc aperture.
Right: $\Delta$PAs for the two samples within the 50 kpc aperture.}
\end{figure*}
\begin{figure*}
\includegraphics[scale=0.75,angle=0]{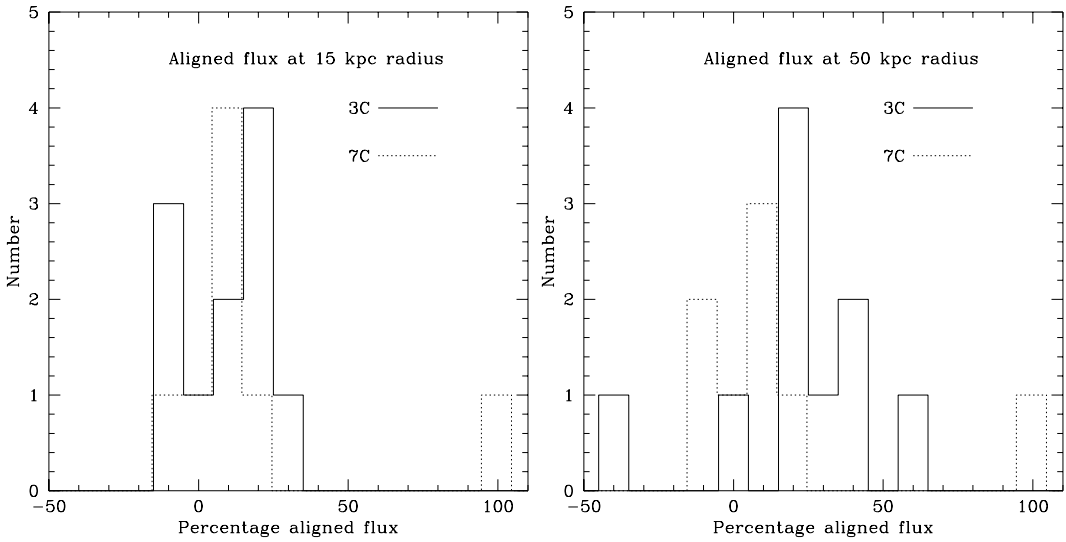}
\caption{Left: Aligned flux percentage
for the 3C and 7C galaxies  within the 15 kpc
aperture.  Right: Aligned flux percentages for the two samples
within the 50 kpc aperture.}
\end{figure*}

\begin{table*}
\caption{Summary of results of statistical tests}
{\small
\begin{tabular}{lllll}
Number & Distributions tested & Type of Test & Result & Probability\\

1 & 7C $\Delta$PA (15 kpc) vs. uniform & K.S. (One-sided) & D = 0.617 & 0.002 \\
2 & 3C $\Delta$PA (15 kpc) vs. uniform & K.S. (One-sided) & D = 0.505 & 0.004 \\
3 & 7C $\Delta$PA (50 kpc) vs. uniform & K.S. (One-sided) & D = 0.361 & 0.193 \\
4 & 3C $\Delta$PA (50 kpc) vs. uniform & K.S. (One-sided) & D = 0.444 & 0.026 \\
& & & & \\
5 & \% Align. Flux (15 kpc): 7C vs. 3C& K.S. (Two-sided) & D = 0.330 & 0.573 \\
6 & \% Align. Flux (50 kpc): 7C vs. 3C& K.S. (Two-sided) & D = 0.550 & 0.095 \\
\
& & & & \\
7 & \% Align. Flux (15 -- 50 kpc): 7C vs. 3C& K.S. (Two-sided) & D = 0.575 & 0.070 \\
8 & \% Align. Flux (15 -- 50 kpc): 7C vs. 3C& K.S. (One-sided) & D = 0.575 & 0.035 \\
\end{tabular}
}
\end{table*}



We also checked the extent to which the two samples are similar in terms of 
properties other than radio luminosity. The spectral index and 
physical size distributions of the two samples 
were compared using the K.S.\ test which showed that the distributions 
of these quantities are very similar; the probability of the null hypothesis
that both are drawn from the same distribution is more than 40 per cent using
a two-sided test. Similarly the magnitudes within 15 and 50 kpc are consistent
with being the same to 30 per cent on a two-sided test. 

\section{Discussion}

\subsection{The comparative strengths and scales of the alignments}

Though the results of Section 5 are admittedly based on small samples and 
are not of great statistical significance, they are quite suggestive.
We seem to have significant alignment on a 15 kpc
scale in a sample of radio galaxies 
whose luminosity is $\sim$20 times fainter than that of the 3C sample.
Of course, even if we are correct in finding almost comparable
percentage alignments in the 7C sample and the 3C sample at $\sim$15 kpc,
this does not necessarily translate into ascribing the same alignment 
mechanism to the two samples. 
An important difference seems to be in the scale of the aligned material.

We have shown that in this sample,
the 3C objects are aligned over a large range in scales, whereas in
the 7C samples the alignments seem mostly to disappear between 15 and 50 kpc.
In the extensively studied $z \sim 1 $ 3C galaxies, of which these are probably
a reasonably typical subset (though at slightly lower $z$), it is known 
that there is a variety of
causes for the alignment effect, some of which would work fairly
well at large scales, such as scattering.
Further evidence for a large-scale aligned component in the 3C radio
galaxies is that we find on average twice as much
flux above the isophotal cutoff in the 3C as in the 7C in the 
15 -- 50 kpc annulus, albeit with great dispersion (although we note that
the difference in the median magnitudes at 15 and 50kpc is similar
in the two samples, suggesting it is relatively compact 
material in the 3C objects which dominates the large-scale alignments).

Though we do not have in our 3C sample any obvious examples in which
the very small scale, ``jet path'' aligned morphology (e.g 3C 324)
dominates, a very small scale alignment mechanism may still 
be contributing, but is swamped by emission from the 
more dominant elliptical host.  We can probably assume that in this 3C
sample the same wide range of mechanisms determined in the $z \sim 1$ 
samples are contributing, and some of these mechanisms 
are obviously effective at scales of $\sim$50 kpc. 

One possibility is that there may simply be less material at 
larger scales in the 7C environments to be aligned or to act as 
a scattering medium. An argument against this is the lack of 
any correlation observed between radio luminosity and galaxy 
cluster environment at $z\sim 0.5$ (Hill \& Lilly 1991). [Our 
NOT data will eventually be used to quantify
the environments of the 7C galaxies directly (Wold, Lacy \& Lilje 1998).]

Why, then, are the galaxies in this 7C sample aligned? 
We have no information, of course, at the very small (0\farcs1) scale 
at which some 3C objects show alignment, and our result, though reasonably
good statistically, is based on a few objects. Furthermore, these
objects themselves
show significant variation in the type of aligned morphology. 
So it is difficult to tell what the likely mechanisms are, and why they
do not seem to be operating at $\sim$ 50 kpc scales. Nevertheless, we
discuss some possibilities below.

\subsection{Mechanisms for alignments at low radio luminosities}

Two of the most common alignment mechanisms known to operate in the
3C sample are scattering and nebular continuum emission. Both of these
should scale approximately as the emission line strengths, which are
found to scale with the radio luminosity to about the 6/7
power (Rawlings \& Saunders 1991; Willott et al.\ 1998). We should therefore
expect contributions from scattering and nebular continuum emission
to be about a factor of 15 less in the 7C sample than in the 3C sample;
were the percentage alignments this much diminished they would have been 
undetectable. Consistent with this is the fact that some of the 7C radio 
galaxies have weak or undetectable emission lines, and the quasar fraction 
is much lower than in 3C (see discussion in Section 5).
One would also expect both optical synchrotron and  
inverse Compton emission (Daly 1992) to scale directly with the radio 
luminosity. 

The probable luminosity dependence of other mechanisms are less 
clear, but they may scale less severely with radio power.
Jet-induced star formation, if it occurs, may even be favoured in lower
luminosity sources where the bulk kinetic energy carried by the radio 
jet is lower. Simulations of shocks propagating into
cool clouds have been carried out by a number of groups. Stone 
\& Norman (1992) and Klein, McKee \& Colella (1994) have looked at the 
purely hydrodynamic effects, and show that shock waves with velocities
appropriate to supernovae (and therefore probably to most radio-jet 
induced shocks too) lead to an adiabatic equation of state in the 
shocked clouds. Such clouds are crushed and shredded due to 
hydrodynamic instabilities.  
Mac Low et al.\ (1994), however, include the effects of 
magnetic fields, and find that an equipartition magnetic field 
may be able to stabilise the clouds against shredding. Foster \& Boss
(1996) include self-gravity in their simulations, and find that  
shocks can promote cloud collapse if the shocked gas remains isothermal
[typically the case for slow ($<100\;$kms$^{-1}$) shocks], but that 
adiabatic shocks still lead to cloud shredding. As jet speeds
are probably relativistic in FRII radio sources with
similar radio luminosities to the 7C sources discussed here  
(Hardcastle et al.\ 1998), it certainly seems 
unlikely that the strong shocks associated with the radio source, for 
example the bow shock, can lead to star formation. Other, 
weaker shocks associated with, for example, backflow or the sideways 
expansion of the lobe, may be more successful, however,
particularly on smaller scales 
within the host galaxy where the densities are high enough that 
some shocks may be close to isothermal. On the basis of these arguments 
it seems unlikely that a strong radio jet will be any more successful at 
forming stars than a weaker one.

The ``selection bias'' effect of Eales (1992) should also operate
effectively in the 7C sample. 
The strength of the effect in a sample with a narrow redshift range is 
dependent on the
luminosity function at the mean redshift of the sample. Fig.\ 6 shows the 
predicted radio -- optical position angle distribution for a variety
of flux limits at $z=0.65$, using the Dunlop \& Peacock (1990) 
luminosity-density evolution model. This shows that the 7C sample
is sufficiently above the break in the luminosity 
function that there is very little dimunition of this effect in the 7C
sample relative to the 3C. 



\begin{figure}
\includegraphics[scale=0.35,angle=-90]{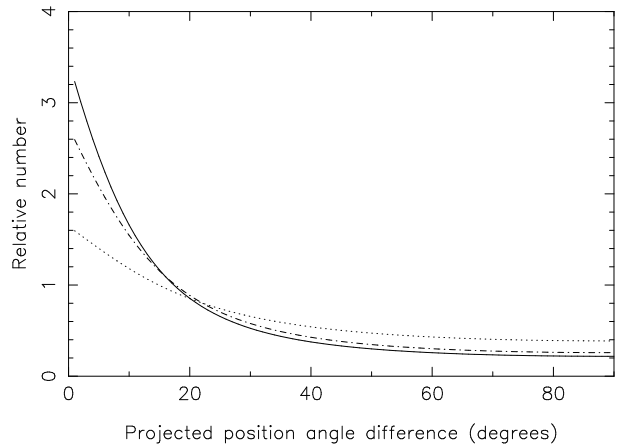}
\caption{The expected distributions in position angle for $z=0.65$
radio sources in the model of Eales (1992) as a function of sample
flux limit, assuming an ellipticity of 0.33 in the density
distribution. Solid curve: 3C ($S_{151}\geq 12$Jy), dot-dash curve:
the 7C sample discussed in this paper ($S_{151}\geq 0.5$ Jy), dotted
curve: a lower flux sample illustrating the fall off in this effect as
the luminosity of the radio sources approaches the break in the 
luminosity function ($S_{151}\geq 0.1$Jy). Although we have only
plotted these curves for one value of the ellipticity, changing the
ellipticity does not significantly affect the luminosity dependence of
this effect.}
\end{figure}



In at least one case (7C 1748+6731) the observed alignment may be due
to a dust disk whose axis aligns with the radio jet axis. Aligned
dust disk axes are reported to be fairly common in the $z<0.5$ 3C 
radio galaxies of the HST snapshot survey (de Koff et al.\ 1995), 
and in other HST data on nearby radio galaxies (van Dokkum \& Franx 1995).
They are thus a viable candidate for producing the 15-kpc scale
alignments in the 7C sample. Only one
of the 7C (and none of the 3C) sample, however, show evidence for dust
lanes, though more may become apparent in the 7C sample at higher
resolution. 


\subsection{Alignments at low redshifts}

Studies of radio-optical alignments in low radio-luminosity radio
galaxies have, with one or two exceptions, 
so far been restricted to low redshifts ($z<0.5$), due to the lack
of suitable radio source samples selected at fainter flux levels than
3C. A few individual cases of alignments have been found,
but no systematic survey has been fully written up in the literature, although 
Dey \& van Breugel (1993) report finding 
aligned UV emission in about 30 per cent of nearby ($0.08<z<0.2$) 
powerful radio galaxies. Since the pool of potential sources is very
large, and only ``interesting'' cases have been published, it is hard
to assess whether the low redshift cases are merely coincidences or
not, and hard to determine whether there is real evolution between
$z\sim 0$ and $z\sim 0.65$. 

One systematic survey that has been made is the HST snapshot
survey. Objects in the same luminosity range as the 7C objects are in
the redshift range $0.1 \stackrel{<}{_{\sim}} z \stackrel{<}{_{\sim}}
0.3$ (Fig.\ 8). 
In Fig.\ 7 we plot the difference between the radio and optical
position angles for 3C objects in this redshift range, 
as tabulated by de Koff et al.\ (1998). Although the isophotal limits and
the rest-frame wavelength of observations are somewhat different, 
both sets of images are taken above the 4000\AA$\;$break, so the underlying
stellar population might be expected to dominate in both
samples. Although there may be a weak alignment trend (note, for
example, the lack of objects in the 80-90$^{\circ}$ bin), there is
clearly no strong alignment effect in this sample.

\begin{figure}
\includegraphics[scale=0.4]{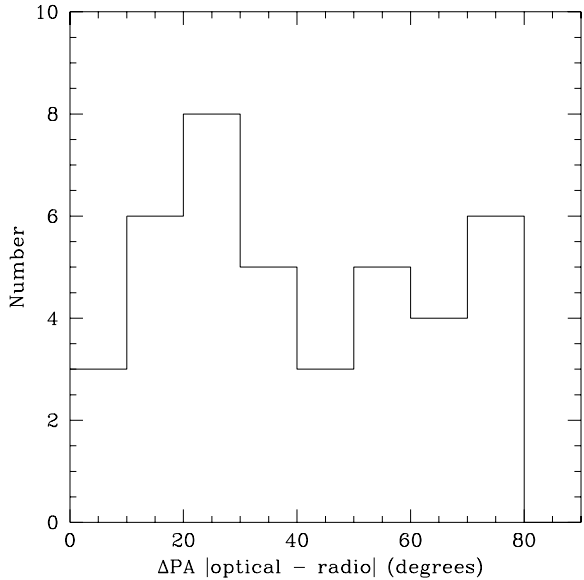}
\caption{The distribution of $\Delta$PA for the 40 objects in the 
HST snapshot survey with $0.1<z<0.3$ (de Koff et al.\ 1998).}
\end{figure}

\begin{figure}
\includegraphics[scale=0.4]{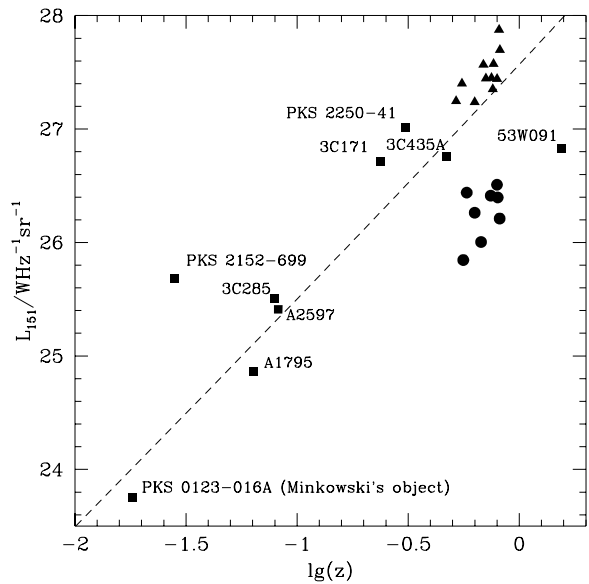}
\caption{A radio luminosity -- redshift plane diagram illustrating the
range in radio luminosities of the radio -- optical alignments
discussed in this paper. Solid triangles: the 3C sample of Table 3, 
solid circles: the 7C sample of Table 1. Solid squares represent other 
sources from the literature. The dashed line corresponds to the flux 
limit of the 3CR from which the HST snapshot survey objects were selected.}
\end{figure}

The individual low redshift cases of radio--optical alignment 
that have been published actually span a wide range in radio
luminosity. To illustrate this, in Fig.\ 8 we 
plot some of the better known low
luminosity aligned objects, together with our 3C and 7C samples, on
the radio-luminosity redshift plane. 

Perhaps the most famous example of an FRI radio source 
with aligned emission is PKS0123-016A, where the blue peculiar 
galaxy known as Minkowski's object is seen apparently interacting 
with the radio jet (van Breugel et al.\ 1985). 
The FRI radio sources in A1795 and A2597 have been studied by McNamara
and collaborators. They find excess blue light in the host galaxies
which appears to trace the radio lobes but which is unpolarised 
(McNamara et al.\ 1997; McNamara et al.\ 1996; Sarazin et al.\ 1995; 
McNamara \& O'Connell 1993). All these cases of the alignment effect
in FRI radio galaxies seem best explained via jet-induced star
formation. Indeed, as discussed in Section 6.2, jet-induced star formation 
may be easier in sources with lower jet velocities.

3C285 is an FRII radio source close to the FRI/FRII boundary which has
a blue aligned component projected $\approx 70$kpc from the 
nucleus which van Breugel \& Dey (1993) ascribe to jet-induced star 
formation (the host galaxy itself is not aligned). 
PKS 2152-699, another low luminosity FRII has an aligned blue
knot whose light is polarised at $90^{\circ}$ to the radio axis, which
suggests that scattered light from the nucleus is responsible. 
(di Serego Alighieri 1989; Fosbury et al.\ 1998). 
Both these objects are a little less radio-luminous than the 7C objects. 

Aligned emission has also been seen in more powerful FRII radio
sources, notably 3C171 (where nebular continuum seems to be
responsible; Clark et al.\ 1998), PKS 2250-41 (Clark et al.\ 1997) 
and 3C435A (Rocca-Volmerange et al.\ 1994). The latter two cases are
discussed below.

\subsection{Red companion alignments}

Two of the objects in the 7C sample, 7C1807+6831 and 7C1826+6510 have 
aligned components with red stellar populations. Furthermore, the morphology
of these components is peculiar, with both having fainter aligned material 
strung out along the line joining them to the radio source host galaxy. 
There are a number of other
similar red aligned objects in the literature, and they provide some of the 
most puzzling manifestations of the alignment effect. Examples include
3C34 (Best et al.\ 1997a), 3C114 (Dunlop \& Peacock 1993; Leyshon \& Eales 
1998) 3C212 (the SE component; Ridgway \& Stockton 1997; Stockton \& Ridgway
1998), 53W091 (Spinrad et al.\ 1997), 3C356 (whichever of `a' or `b' is the
true host; Lacy \& Rawlings 1994; Cimatti et al.\ 1997), 
3C435A (Rocca-Volmerange et al.\ 1994), PKS 2250-41 (Clark
et al.\ 1997)  and 3C441 (Lacy et al.\ 1998). 
These objects can be divided into two classes, those where the companion is 
just beyond a radio hotspot (3C212, 53W091, 3C435A, 3C441, PKS2250-41) 
and those where it is within a lobe
(3C34, 3C114, 3C356, 7C 1807+6831, 7C1826+6510).
 
Those with optical components close to hotspots can 
probably be explained in terms of a radio jet 
colliding with a galaxy in the same group or cluster as the host. The
increase in external density will raise the ram pressure in the head
of the source, enhancing the synchrotron emissivity and thereby 
pushing the source into a flux-limited sample. 

Those with optical components in the lobe are more puzzling [except perhaps 
for 3C114 which is polarized in $K$-band; Leyshon \& Eales (1998)]. 
Chance, or unusually well-aligned
manifestations of the selection effect of Eales (1992) could explain 
their presence, but if so their often peculiar morphogies 
(e.g.\ 3C34 and the two 7C objects in this paper) remain a puzzle.
A possible, but perhaps unlikely explanation, given the discussion in Section 
6.2, is that jet-induced star formation could occur with an IMF
weighted heavily towards massive stars and lead to an aligned
component dominated by a red stellar population in $\sim 10^7$ yr 
(Lacy \& Rawlings 1994; Best et al.\ 1997a).

\subsection{Implications for photometry of radio galaxy hosts}

Could a component of aligned light on large scales whose luminosity is 
proportional to that of the AGN affect near-infrared studies of $z\sim 1$ 
radio galaxies? Roche, Eales \& Rawlings (1998) find
that the $K$-band sizes and magnitudes of $z\sim 1$ radio source hosts differ
significantly between the 3C and 6C samples, which are only a factor
of four different in radio flux limit, and Best et al.\ (1998) claim to 
find extended haloes around their HST images of 3C radio 
galaxies which they suggest may be cD galaxy haloes. However, there is 
a significant $K$-band alignment effect seen in 3C radio galaxies 
(Rigler et al.\ 1992; Dunlop \& Peacock 1993; Ridgway \& Stockton 1997)
which is not present in the 6C objects (Eales et al.\ (1997). 
Also, the 6C and 7C samples follow similar 
$K-z$ relations despite the factor of five between the 6C and 7C flux limits
(Rawlings et al.\ 1997), again consistent with the idea 
that the 3C galaxies are significantly contaminated by emission which scales
with the power of the AGN. Estimates of the amount of AGN light 
contributing in the $K$-band fall short of the amount required to explain
the difference between the 3C and 6C/7C $K-z$ relations, however.
Rigler et al.\ (1992) argue that the aligned component seen in the
3C galaxies contributes only $\approx 10$ per cent of the total 
near-infrared emission, and Simpson, Rawlings \& Lacy (1998) 
show that the effect of the reddened quasar nucleus is only about 
a 15 per cent correction to the mean magnitude at $z\approx 1$. This 
would predict that 3C galaxies should be about 0.25 mag brighter than 6C 
ones, compared to the 0.59 mag difference 
between the 3C and 6C mean galaxy magnitudes (in 63.9 kpc apertures)
measured by Eales et al.\ (1997). As yet there is also no direct evidence that 
the larger scale-sizes of the 3C radio galaxies are connected to the 
aligned light. Nevertheless, the difference in the mean magnitudes of 3C and 
6C galaxies after allowing for the contributions of aligned light and 
reddened quasar nuclei is sufficiently small that high spatial resolution 
infrared imaging and spectroscopy will probably be required to finally resolve
this issue. 

\section{Conclusions}

A possible explanation for the scale difference we see in the
alignments of the 3C and 7C radio galaxies is that the 
alignment mechanisms that operate at the large scales in the 3C are the 
most luminosity dependent. Both scattering and nebular continuum could
be argued to fall into this catagory provided enough material exists at large
radius to intercept light from the hidden quasar. 

The small-scale alignment that we seem to see 
in the 7C sample indicates that some less luminosity-dependent alignment
mechanism, effective at moderately small scales,
may be contributing to the ``alignment effect'' that we see in radio
sources. That the 3C alignments extend to larger scales could mean
that this less-luminosity dependent alignment mechanism is swamped
by other effects, such as scattering, in the more powerful radio
sources. Possible candidates for this small-scale alignment mechanism include
jet-induced star formation, dust discs with axes parallel to the radio
axis, and the selection effect of Eales (1992) (assuming the PA
of the host galaxy is parallel to that of the major axis of the density 
distribution on the scale size of the radio source). 

In addition, there are two cases of more extended aligned light in
the 7C sample. In both cases old stellar populations are seen in the 
brightest aligned components. They may be present due to selection
effects, but jet-induced star formation with a peculiar IMF cannot
be ruled out, and may explain the unusual morphology of the aligned light.

The detection of alignment in the $z\sim 0.7$ 7C sample also suggests the
alignment effect has a component which is truly due to cosmic
evolution in the host galaxies of radio sources, and not just 
due to the increase in radio luminosity with
redshift present in flux-limited samples. Systematic observations of low
redshift 3C radio galaxies in directly comparable wavebands are required
to confirm this, however.

The presence of an alignment mechanism (or mechanisms) 
which scales only weakly with
radio luminosity may imply that even a fairly weak radio AGN can
influence the morphology and perhaps even stellar content of its host. 
This would have implications for attempts to search for old elliptical
galaxies at high redshifts through spectroscopy of faint
radio sources (e.g.\ Dunlop et al.\ 1996). 
On the other hand, if dust lanes or selection effects 
are responsible, then
such galaxies may be ideal for studying the oldest stellar populations
at high redshift. It is thus important to 
establish what this small-scale alignment 
mechanism is, and it is clear that higher resolution 
observations of a fainter radio source sample than 3C would be of
great help in this respect. 

\section*{acknowledgments}

We thank Christian Kaiser for helpful comments on the manuscript and 
the staff at the NOT and WHT for their help with the
observations. SER is supported by a PPARC PDRA grant.
The NOT is operated jointly by Denmark, Finland, Iceland, Norway 
and Sweden, and the WHT by the Isaac Newton Group both on the Island 
of La Palma in the Spanish Observatorio del Roque de los Muchachos 
of the Instituto de Astrofisica de Canarias. This paper was 
also partly based on observations made 
with the NASA/ESA Hubble Space Telescope, obtained from the data archive at 
the Space Telescope Science Institute. STScI is operated by the Association of
Universities for Research in Astronomy, Inc. under the NASA contract 
NAS 5-26555.

\end{document}